\documentclass[a4 paper]{article}
\usepackage{amssymb}
\usepackage{amsfonts}
\usepackage{graphicx}
\usepackage{pdfpages}
\usepackage{setspace}
\usepackage{multirow}
\usepackage{amsmath}
\usepackage{array}
\usepackage{subfigure}
\usepackage{xcolor}
\usepackage{mathrsfs}
\usepackage{booktabs}
\usepackage{bbding}
\usepackage{bbding}
\usepackage{verbatim}
\usepackage{bbding}

\usepackage{INTERSPEECH2022}

\ninept
\makeatletter
\renewcommand\normalsize{%
	\abovedisplayskip 3\p@ \@plus3\p@ \@minus3\p@
	\abovedisplayshortskip \z@ \@plus3\p@
	\belowdisplayshortskip 6\p@ \@plus3\p@ \@minus3\p@
	\belowdisplayskip \abovedisplayskip
	\let\@listi\@listI}
\makeatother

\title{TaylorBeamformer: Learning All-Neural Beamformer for Multi-Channel Speech Enhancement from Taylor's Approximation Theory}
\name{Andong Li$^{1, 2}$, Guochen Yu$^{1}$, Chengshi Zheng$^{1,2}$, Xiaodong Li$^{1,2}$}
\address{
	$^1$Key Laboratory of Noise and Vibration Research, Institute of Acoustics, Chinese Academy of Sciences, Beijing, China\\
	$^2$University of Chinese Academy of Sciences, Beijing, China
}
\email{ \{liandong, cszheng, lxd\}@mail.ioa.ac.cn, \{yuguochen\}@cuc.edu.cn}
\begin{document}
\maketitle
\begin{abstract}
While existing end-to-end beamformers achieve impressive performance in various front-end speech processing tasks, they usually encapsulate the whole process into a black box and thus lack adequate interpretability. As an attempt to fill the blank, we propose a novel neural beamformer inspired by Taylor's approximation theory called TaylorBeamformer for multi-channel speech enhancement. The core idea is that the recovery process can be formulated as the spatial filtering in the neighborhood of the input mixture. Based on that, we decompose it into the superimposition of the 0th-order non-derivative and high-order derivative terms, where the former serves as the spatial filter and the latter is viewed as the residual noise canceller to further improve the speech quality. To enable end-to-end training, we replace the derivative operations with trainable networks and thus can learn from training data. Extensive experiments are conducted on the synthesized dataset based on LibriSpeech and results show that the proposed approach performs favorably against the previous advanced baselines.

\end{abstract}
\vspace{-0.2cm}
\section{Introduction}
\label{sec:introduction}
\vspace{-0.10cm}
Multi-channel speech enhancement (MCSE) aims at extracting target speech from multiple noisy-reverberant microphone recording signals. A handful of traditional beamforming-based algorithms have been proposed in the past decades~{\cite{gannot2017consolidated}}. Recently, with the proliferation of deep neural networks (DNNs), neural beamformers are viewed as a type of promising technique in various applications like far-field speech restoration and automatic speech recognition (ASR)~{\cite{heymann2015blstm, qian2018deep}}.

As a typical scheme, a DNN is first utilized to extract the target speech, followed by traditional spatial filters~{\cite{erdogan2016improved}}, \emph{e.g.}, minimum variance distortionless response (MVDR), and multi-channel Wiener filter (MWF). The downside of this paradigm is that the two parts are processed separately and the performance of spatial filtering utterly hinges on the pre-estimation of the previous stage. Besides, the performance may suffer from heavy performance degradation under frame-level processing conditions. Another class regards MCSE as the extension of the single-channel case, where the spatial cue is usually extracted either manually or implicitly as the auxiliary feature to assist the single-channel SE system~{\cite{gu2019neural, wang2018combining}}. Despite the efficacy, it abandons the preponderance of spatial filtering, leading to heavy nonlinear distortion under adverse acoustic scenarios.

More recently, all-neural frame-level beamformers are investigated in either time domain~{\cite{luo2020end, luo2021implicit, ochiai2020beam}} or time-frequency (T-F) domain~{\cite{zhang2021adl, halimeh2021complex, li2021embedding, casebeer2021nice}}. Among these methods, frame-wise beamformer weights are estimated by the network to implement the beamforming process and yield impressive performance in noise suppression and source separation tasks. Despite the effectiveness, they often neglect the reverberation components and only consider the processing of directional sound sources. This is because late reverberation is usually assumed to be spatially diffused~{\cite{schwartz2014multi}} and it is thus difficult to be canceled with the beamforming technique, especially when the number of microphones is limited. Besides, as the network is trained in an end-to-end manner, it is expected to extract both spatial and spectral cues and serve as the  spatial-spectral processor~{\cite{li2021embedding, tan2022neural}}. Therefore, the whole process is actually encapsulated into a black box and lacks internal mechanism and adequate interpretability.

According to the Woodbury matrix identity, MWF can be decomposed into the tandem format of MVDR and spectral post-processing. Similarly, we would like to raise a question, \emph{whether it is possible to decouple spatial and spectral processing modes for all-neural frame-wise beamformer system?} As a response, we propose an all-neural beamformer framework termed \textbf{TaylorBeamformer} to address simultaneous denoising and dereverberation. To be specific, we rethink the beamforming process and formulate the target extraction into the distortionless spatial filtering in the neighboring point of noisy input. Our core insight is that if we are able to cancel the interference component within the input in prior, the target speech can be perfectly recovered via spatial filtering theoretically. This process can be represented via Taylor's approximation theory and we can thus formulate the whole beamforming process into the superimposition of the 0th-order non-derivative and high-order derivative terms, where the former implements the spatial filtering to suppress the directional noise and the latter is tasked with residual interference cancellation. Furthermore, we propose to replace the derivative terms with learnable modules to support end-to-end training. To our best knowledge, it is the first time to apply Taylor's approximation theory into the speech front-end field, which provides a deep insight into all-neural beamformers and can help us understand the logic behind the model.

The rest of the paper is structured as follows. In Section~{\ref{sec:problem-formulation}}, the problem formulation is introduced. In Section~{\ref{sec:proposed-approach}}, the proposed method is presented in detail. Section~{\ref{sec:experiments-setup}} gives the experimental setup, and results and analysis are presented in Section~{\ref{sec:results-and-analysis}}. Some conclusions are drawn in Section~{\ref{sec:conclusion}}.  
\vspace{-0.40cm}
\begin{figure*}[t]
	\centering
	\centerline{\includegraphics[width=1.65\columnwidth]{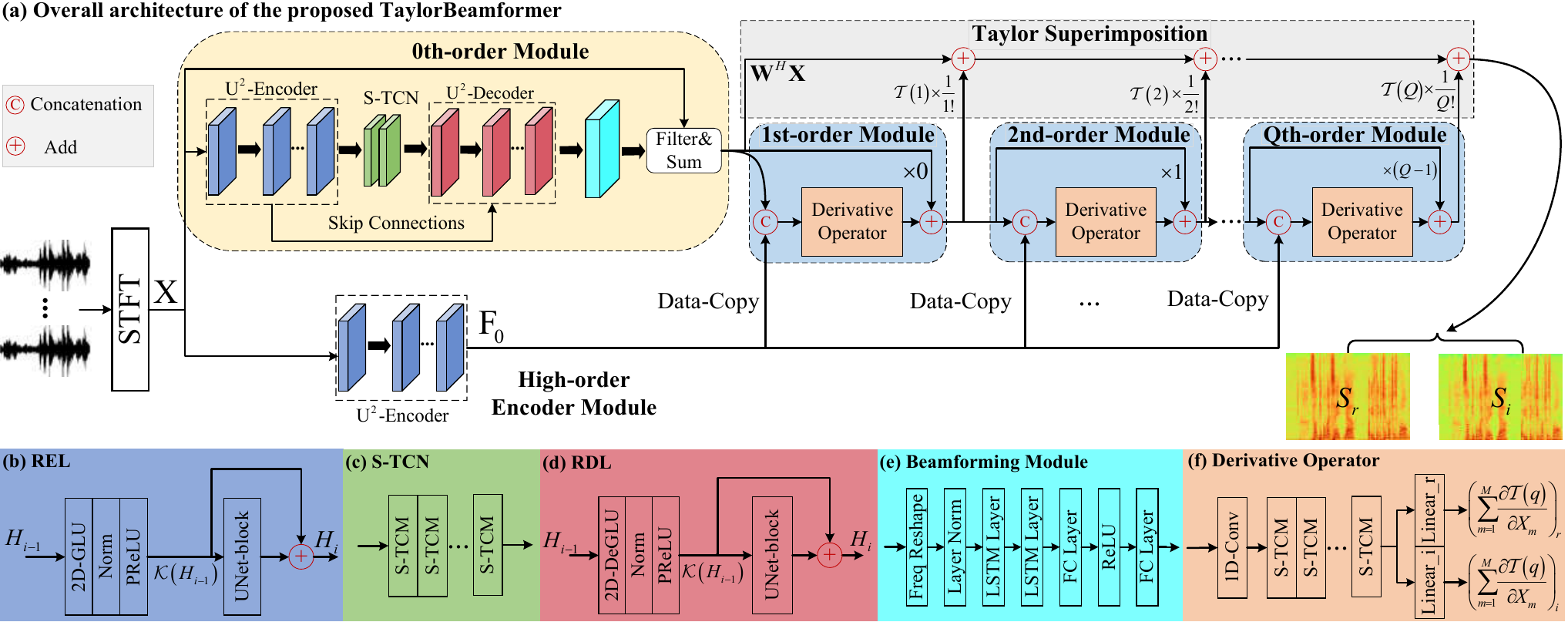}}
	\caption{(a) Overall diagram of the proposed TaylorBeamformer. (b) Internal structure of recalibration encoding layer (REL). (c) Internal structure of S-TCN. (d) Internal structure of recalibration decoding layer (RDL). (e) Internal structure of beamforming module. (f) Internal structure of derivator operator. Different modules are highlighted with different colors for better visualization.}
	\label{fig:architecture}
	\vspace{-0.6cm}
\end{figure*}
\section{Problem formulation}
\label{sec:problem-formulation}
\vspace{-0.20cm}
With short-time Fourier transform (STFT), the noisy-reverberant signals observed by an $M$-channel array can be modeled as
\begin{gather}
\label{eqn1}
\mathbf{X}_{t, f} = \mathbf{c}_{f}S_{t, f} + \mathbf{V}_{t, f} + \mathbf{N}_{t, f} = \mathbf{S}_{t, f} + \mathbf{V}_{t, f} + \mathbf{N}_{t, f} , 
\end{gather}
where $\left\{\mathbf{X}_{t, f}, \mathbf{S}_{t, f}, \mathbf{V}_{t, f}, \mathbf{N}_{t, f}\right\}\in\mathbb{C}^{M\times1}$ denote the mixture, anechoic speech, reverberant speech, and reverberant noise at the time index of $t\in\left\{1,\cdots,T\right\}$ and the frequency index of $f\in\left\{1,\cdots,F\right\}$, respectively. $\mathbf{c}_{f}\in\mathbb{C}^{M\times1}$ denotes the relative transfer function (RTF) of the speech. Without loss of generality, the first microphone is picked as the reference channel and sound sources are assumed to be static within each utterance.

Different from previous literature where only directional noise is removed~{\cite{luo2020end, zhang2021adl, li2021embedding}}, we regard both reverberation and noise components as the interference and attempt to remove it with the beamforming method and recover anechoic speech:
\begin{gather}
\label{eqn2}
\widetilde{S}_{t, f} = \mathbf{W}_{t, f}^{H}\mathbf{X}_{t, f},
\end{gather}
where $\mathbf{W}_{t, f}\in\mathbb{C}^{M\times1}$, and $(.)^{H}$ denotes the conjugate transpose. Note that the weights are time-dependent as frame-level beamformer is adopted. Now onwards, we will drop the T-F subscript $\left\{t, f\right\}$ if no confusion arises, and if we indicate the tensor, we will write it with uppercase letters. Take the $m$th channel as an example, after filtering, the filtered signal can be given by
\begin{gather}
\label{eqn3}
W_{m}^{*}X_{m} = W_{m}^{*}\left(S_{m} + R_{m}\right),
\end{gather}
where $(.)^{*}$ denotes the conjugate operation, subscript $m$ denotes the channel index, and $R_{m} = V_{m} + N_{m}$ denotes the interference. If the distortionless spatial filter like MVDR is adopted, we can rewrite Eqn.{(\ref{eqn3})} by summing all the channels as
\begin{gather}
\label{eqn4}
\sum_{m=1}^{M}W_{m}^{*}X_{m} = S + \sum_{m=1}^{M}W_{m}^{*}R_{m}.
\end{gather}

From Eqn.({\ref{eqn4}), one can see that despite MVDR can guarantee low distortion of target speech, some residual interference components still remain, which need to be further removed by postprocessing. Assuming there exists a prior term $\delta_{m}$ to cancel the interference in advance, Eqns.~{(\ref{eqn3})-(\ref{eqn4})} can be rewritten as
\begin{gather}
\label{eqn5}
W_{m}^{*}\left(X_{m} + \delta_{m}\right) = W_{m}^{*}S_{m},\\
\sum_{m=1}^{M}W_{m}^{*}\left(X_{m}+\delta_{m}\right) = \sum_{m=1}^{M}W_{m}^{*}S_{m} = S,
\end{gather}
where $\delta_{m} = -R_{m}$. Furthermore, let us denote $G_{m}\left(X\right) = W_{m}^{*}X$, Eqn.(6) can be abstracted into a more general case:
\begin{gather}
\label{eqn6}
S = \sum_{m=1}^{M}G_{m}\left(X_{m} + \delta_{m}\right).
\end{gather}

Eqn.({\ref{eqn6}}) implies that for each microphone channel, if we can access $X_{m}+\delta_{m}$, the neighboring point of $X_{m}$, in advance, then we can perfectly recover the target speech by MVDR theoretically. However, the term $\delta_{m}$ tends to be unknown in practical scenarios. In this regrad, if the above function is differentiable to every order, we can resolve it with infinite Taylor's series expansion at $X_{m}$ as
\begin{gather}
\label{eqn7}
S = \sum_{m=1}^{M}G_{m}\left(X_{m}\right) + \sum_{q=1}^{+\infty}\frac{1}{q!}\sum_{m=1}^{M}\frac{\partial^{q}G_{m}\left(X_{m}\right)}{\partial^{q}X_{m}}\delta_{m}^q,
\end{gather}
where the 0th-order term can be rewritten as the spatial fitering:
\begin{equation}
\label{eqn8}
\sum_{m=1}^{M}G_{m}\left(X_{m}\right) = \sum_{m=1}^{M}W_{m}^{*}X_{m} = \mathbf{W}^{H}\mathbf{X}.
\end{equation}

As such, we provide a disentanglement perspective toward the recovery process. Concretely, the 0th-order term serves as the spatial filtering toward the noisy mixture to remove most of the directional interference while guaranteeing the low distortion of target speech. For high-order terms, they work as the spectral canceller to further suppress the residual noise by superimposition. Recap that MWF can also be decomposed into the concatenation of spatial and spectral modules. The major difference is that for MWF, the spectral part is a simple single-channel postprocessing while we adopt the superimposition of various high-order derivative terms to accomplish this step.
\vspace{-0.35cm}
\section{Proposed approach}
\label{sec:proposed-approach}
\vspace{-0.2cm}
\subsection{Correlation between adjacent high-order terms}
\label{sec:correlation-between-adjacent-terms}
\vspace{-0.2cm}
For practical implementation, the total number of orders is set to $Q$. To adapt the Taylor series into end-to-end training, it is imperative to derive the relation between adjacent high-order terms. As such, let us notate the $q$th order term as
\begin{equation}
\label{eqn9}
\mathcal{T}\left(q\right) = \sum_{m=1}^{M}\frac{\partial^{q}G_{m}\left(X_{m}\right)}{\partial^{q}X_{m}}\delta_{m}^{q},
\end{equation}
where the factorial term is neglected for derivation convenience. For Eqn.({\ref{eqn9}}), we differentiate $\mathcal{T}\left(q\right)$ with respect to $X_{m}$:
\begin{equation}
\label{eqn10}
\frac{\partial \mathcal{T}\left(q\right)}{\partial X_{m}} = \frac{\partial}{\partial X_{m}}\left(  \frac{\partial^{q}G_{m}\left(X_{m}\right)}{\partial^{q}X_{m}}\right)\delta_{m}^{q} + \frac{\partial^{q}G_{m}\left(X_{m}\right)}{\partial^{q}X_{m}} \frac{\partial}{X_{m}}\delta_{m}^{q}.
\end{equation}

Considering
\begin{equation}
\label{eqn11}
\frac{\partial}{\partial X_{m}}\left(  \frac{\partial^{q}G_{m}\left(X_{m}\right)}{\partial^{q}X_{m}}\right) = \frac{\partial^{q+1}G_{m}\left(X_{m}\right)}{\partial^{q+1}X_{m}},
\end{equation}
\begin{equation}
\label{eqn12}
\frac{\partial}{X_{m}}\delta_{m}^{q} = \frac{\partial\delta_{m}^{q}}{\delta_{m}}\frac{\delta_{m}}{X_{m}} = -q\delta_{m}^{(q-1)}.
\end{equation}

Substituting Eqns.~({\ref{eqn11}})-({\ref{eqn12}}) into Eqn.({\ref{eqn10}}), then multiplying $\delta_{m}$ on both sides and summing all the channels, we can derive the following recursive formula between adjacent order terms:
\begin{equation}
\label{eqn13}
\mathcal{T}(q+1) = q\mathcal{T}(q) + \sum_{m=1}^{M}\frac{\partial\mathcal{T}\left(q\right)}{\partial X_{m}}.
\end{equation}

It is evident that the second derivative term on the right side of the above equation seems too difficult to calculate in practical implementation. As such, we propose to replace it with a trainable network and can thus learn the mathematical representation from training data. In the experimental section, we find that by end-to-end training, high-order modules can cancel the residual interference and further improve the speech quality.
\vspace{-0.35cm}
\subsection{System forward stream}
\label{sec:system-forward-stream}
\vspace{-0.2cm}
To simulate the structure of Taylor's series expansion, we elaborately devise a framework to realize this process, whose overall diagram is shown in Figure~{\ref{fig:architecture}}(a). There are three major parts, namely the 0th-order module, high-order encoder module, and multiple high-order modules. In the 0th-order module, the network aims to simulate the behavior of frame-level spatial filtering to suppress directional interference and preserve target speech. As the derivative term in Eqn.({\ref{eqn13}) involves the original inputs, the high-order encoder module is employed as the feature extractor to guide the modeling of high-order parts. For high-order modules, following the recursive formula in Eqn.({\ref{eqn13}), we iteratively update the output of each high-order term and then superimpose all of them to output the final estimation. The overall forward stream is formulated as
\begin{gather}
\label{eqn14}
\mathbf{W} = \mathcal{F}_{bf}\left(\mathbf{X}\right),\\
\widetilde{\mathbf{S}}_{0} = \mathbf{W}^{H}\mathbf{X},\\
\mathbf{F}_{0} =\mathcal{F}_{en}\left(\mathbf{X}\right),\\
\mathcal{T}\left(q+1\right) = q\odot\mathcal{T}\left(q\right) + \mathcal{F}_{deri}\left(Cat\left( \mathbf{F}_{0}, \mathcal{T}\left(q\right)   \right)\right),\\
\widetilde{\mathbf{S}} = \widetilde{\mathbf{S}}_{0} + \sum_{q=1}^{Q}\frac{1}{q!}\mathcal{T}\left(q\right),
\end{gather}
where $\left\{ \mathcal{F}_{bf}, \mathcal{F}_{en}, \mathcal{F}_{deri} \right\}$ denote the functions of the 0th-order, high-order encoder, and derivator operator. $\odot$ is the element-wise multiplier, and $Cat$ denotes the concatenation operation.
\vspace{-0.4cm}
\renewcommand\arraystretch{0.92}
\begin{table}[t]
	\setcounter{table}{0}
	\caption{Ablation study on the proposed TaylorBeamformer. The values are averaged among the test set. \textbf{BOLD} indicates the best score in each case.}
	\large
	\centering
	\resizebox{\columnwidth}{!}{
		\begin{tabular}{c|ccccccc}
			\specialrule{0.1em}{0.25pt}{0.25pt}
			\multirow{2}*{Entry} &\multirow{2}*{$Q$} &Param. &MACs   &\multirow{2}*{PESQ$\uparrow$} &\multirow{2}*{ESTOI(\%)$\uparrow$} &\multirow{2}*{SISDR(dB)$\uparrow$} &\multirow{2}*{DNSMOS$\uparrow$}\\
			& &(M) &(G/s) & & & &\\
			\specialrule{0.1em}{0.25pt}{0.25pt}
			1a &0 &\textbf{2.36} &\textbf{6.64} &2.44 &66.13 &3.86 &2.89 \\
			1b &1 &3.95 &8.45  &2.74 &72.99 &5.11 &3.19\\
			1c &2 &4.77 &8.54 &2.78 &74.24 &5.52 &3.21\\
			1d &3 &5.60 &8.62 &2.80 &74.75 &5.91 &3.25\\
			1e &4 &6.42 &8.70  &2.79 &74.68 &5.84 &3.25\\
			1f &5 &7.25 &8.79 &\textbf{2.84} &\textbf{75.43} &\textbf{6.15} &\textbf{3.26}\\
			1g &6 &8.07 &8.87 &2.82 &75.27 &6.12 &\textbf{3.26}\\
			\specialrule{0.1em}{0.25pt}{0.25pt}
			2a &3 &5.58 &8.51 &2.29 &59.73 &1.94 &3.08\\
			2b &3 &5.59 &8.57 &2.74 &72.86 &5.36 &3.22\\
			\specialrule{0.1em}{0.25pt}{0.25pt}
	\end{tabular}}
	\label{tbl:ablation-studies}
	\vspace{-0.6cm}
\end{table}
\vspace{-0.4cm}
\subsection{Network structure}
\label{sec:network-structure}
\vspace{-0.2cm}
Any existing modules can be chosen to adapt to the proposed framework. In this study, we adopt the similar network structure of our previous work~{\cite{li2021embedding}}. To be specific, in the 0th-order module, the ``Encoder-TCNs-Decoder'' structure is adopted to extract both spatial-spectral features, where the UNet-block is inserted after the 2D-(De)GLU to further recalibrate the feature distribution (see Figure~{\ref{fig:architecture}}(b)(d))~{\cite{qin2020u2}}. For sequence modeling, the squeezed version of temporal convolution networks called S-TCN~{\cite{li2021two}} is adopted, where multiple S-TCMs are stacked to gradually enlarge the temporal receptive field, as shown in Figure~{\ref{fig:architecture}}(c). After the decoder, we generate a 3D tensor, which is expected to incorporate both spectral and spatial discriminative cues. Then the beamforming module is devised to simulate the behavior of traditional frequency-wise beamformers and generate the filter weights, as shown in Figure~{\ref{fig:architecture}}(e). For each derivative operator, similar to S-TCN, multiple S-TCMs are concatenated, and two linear layers are adopted as the network output to generate the real and imaginary (RI) parts of the derivative term, as shown in Figure~{\ref{fig:architecture}}(f). Due to the space limit, we may refer the readers to~{\cite{li2021embedding}} for detailed network introduction.
\vspace{-0.4cm}
\subsection{Loss function}
\label{sec:loss-function}
\vspace{-0.2cm}
To enforce the 0th-order and high-order modules work as the spatial filter and residual canceller, respectively, weighted multi-objective loss is adopted. Specifically, oracle time-invariant MVDR (TI-MVDR) is applied to generate the beamforming output, which serves as the intermediate label of the 0th-order module. Besides, we supervise the final output with anechoic target speech. The loss function can be given as
\begin{equation}
\label{eqn15}
\mathcal{L} = \alpha\mathcal{L}_{bf} + \beta\mathcal{L}_{sp},
\end{equation}
where $\mathcal{L}_{bf}$, $\mathcal{L}_{sp}$ denote the loss functions of spatial filtering and target spectrum recovery. $\left\{\alpha, \beta\right\}$ are the weighted coefficients, which are set to 1 empirically. Similar to~{\cite{li2021two}}, RI loss together with the magnitude constraint is adopted for each loss term.
\vspace{-0.40cm}
\renewcommand\arraystretch{0.80}
\begin{table*}[t]
	\setcounter{table}{1}
	\caption{Results comparison with advanced baselines. The values are specified with PESQ/ESTOI/SISDR/DNSMOS formats. ``Cau.'' denotes whether the system is implemented with causal setting.}
	\normalsize
	\setlength{\tabcolsep}{3pt}
	\centering
	\resizebox{0.90\textwidth}{!}{
		\begin{tabular}{c|ccc|ccccc}
			\hline
			\multirow{2}*{Systems} &Para. &MACs
			&\multirow{2}*{Cau.}  &\multicolumn{4}{c}{Angle between target speech and noise} &\multirow{2}*{Average}\\
			\cline{5-8}
			&(M) &(G/s) & &$0-15^{\circ}$ &$15^{\circ}-45^{\circ}$ &$45^{\circ}-90^{\circ}$ &$90^{\circ}-180^{\circ}$ & \\
			\cline{1-9}
			Noisy  &- &- &- &1.60/40.45/-5.67/2.49 &1.59/40.06/-5.83/2.51 &1.61/40.61/-5.77/2.49 &1.63/40.12/-5.78/2.50 &1.61/53.22/-5.76/2.50 \\
			\hline
			MMUB &\textbf{1.97} &\textbf{4.09} &\Checkmark &1.87/51.73/1.01/2.72 &2.11/57.87/2.11/2.87 &2.29/61.11/3.25/2.89 &2.37/64.04/4.25/2.96 &2.16/58.69/2.65/2.86\\
			EaBNet &2.84 &7.38 &\Checkmark &2.44/66.93/3.61/3.08 &2.67/71.62/4.22/3.22 &2.81/73.35/5.24/3.23 &2.90/75.77/6.17/3.27 &2.70/71.92/4.81/3.20 \\
			Oracle Frame-MVDR &- &- &\Checkmark &2.38/68.54/3.41/2.85  &2.50/70.48/3.84/2.96  &2.59/70.49/4.11/2.98  &2.61/71.94/4.65/2.98  &2.52/70.36/4.00/2.94 \\
			\textbf{TaylorBeamformer} &5.60 &8.62 &\Checkmark &\textbf{2.59/71.15/4.87/3.13 } &\textbf{2.78/74.86/5.53/3.28}  &\textbf{2.88/75.36/6.11/3.28}  &\textbf{2.96/77.62/7.12/3.32}  &\textbf{2.80/74.75/5.91/3.25}\\
			\hline
			FasNet-TAC &2.77 &5.25 &\XSolidBrush  &2.16/57.58/3.71/2.71  &2.32/61.17/3.96/2.82  &2.44/63.91/4.96/2.88 &2.53/66.33/5.70/2.90 &2.36/62.25/4.58/2.83 \\
			EaBNet &2.84 &7.38 &\XSolidBrush &2.71/72.38/4.77/\textbf{3.20} &2.86/75.49/5.15/3.29 &2.97/76.95/6.06/3.31 &3.05/79.07/6.92/3.35 &2.90/75.97/5.73/\textbf{3.29} \\
			Oracle TI-MVDR &- &- &\XSolidBrush &2.43/68.22/8.06/2.85 &2.50/69.92/7.92/2.92  &2.56/69.49/8.23/2.93 &2.58/71.26/8.62/2.94  &2.52/69.73/8.21/2.91 \\
			Oracle TI-MWF &- &- &\XSolidBrush  &2.51/68.80/\textbf{9.47}/2.89  &2.56/70.43/\textbf{9.78}/2.94  &2.64/69.93/\textbf{10.47}/2.98  &2.66/71.81/\textbf{11.59}/2.97  &2.59/70.24/\textbf{10.33}/2.95 \\
			\textbf{TaylorBeamformer} &5.60 &8.62 &\XSolidBrush &\textbf{2.76/74.29/}5.92/3.17 &\textbf{2.93/77.52/}6.43\textbf{/3.31}  &\textbf{3.01/78.22/}6.93\textbf{/3.32}  &\textbf{3.09/80.21/}7.72\textbf{/3.37}  &\textbf{2.95/77.56/}6.75\textbf{/3.29}\\
			\hline
	\end{tabular}}
	\label{tbl:results-comparison}
	\vspace{-0.5cm}
\end{table*}
\section{Experimental setup}
\label{sec:experiments-setup}
\vspace{-0.2cm}
\subsection{Dataset}
\label{sec:dataset}
%\footnote{Here SNR is defined as the ratio between reverberant speech and noise, but all the reverberation components are viewed as the interference and will be removed.}
\vspace{-0.2cm}
We synthesize the multi-channel noisy-clean pairs based on open-sourced LibriSpeech ASR corpus~{\cite{panayotov2015librispeech}}, where \emph{train-clean-100}, \emph{dev-clean}, and \emph{test-clean} are leveraged for training, validation, and model evaluation, respectively. For noise set, around 20,000 types of noises are randomly chosen from the DNS-Challenge corpus\footnote{github.com/microsoft/DNS-Challenge/tree/master/datasets} for training, whose duration is around 55 hours. Multi-channel RIRs are generated with image method~{\cite{allen1979image}} based on a uniform linear array (ULA) with 6 microphones, where the distance between adjacent microphones is 5cm. To generalize well toward different room configurations, the room size randomly changes from 5m-5m-3m to 10m-10m-4m (length-width-height), and the reverberation time (T$_{60}$) ranges from of 0.1s to 0.7s. To adapt the trained model to more general acoustic scenarios, the distance from target/noise source to the microphone varies from 0.5m to 5.0m with 0.5m intervals. The direction of arrival (DOA) difference between target and noise is at least 5$^\circ$. The signal-to-noise ratio (SNR) is randomly selected from $\left[-6\rm{dB}, 6\rm{dB}\right]$. In total, we generate 40,000, 4000 pairs for training and validation. For model evaluations, around 50 types of unseen environmental noises are selected from MUSAN~{\cite{snyder2015musan}}, and four DOA-difference cases (0-15$^{\circ}$, 15$^{\circ}$-45$^{\circ}$, 45$^{\circ}$-90$^{\circ}$, 90$^{\circ}$-180$^{\circ}$) are set, each of which contains 200 testing pairs.
\vspace{-0.35cm}
\subsection{Configurations}
\label{sec:configurations}
\vspace{-0.2cm}
In the 0th-order module, the kernel size of 2D-GLU and UNet-block are set to (1, 3) and (2, 3), respectively, and the number of channels is set to 64 by default. The hidden node in the beamforming module remains 64, and the output channel dimension is $2\times6 = 12$ for spatial filter weights generation. For both S-TCN and derivative operators, two groups of S-TCMs are adopted, each of which contains 4 S-TCMs and the kernel size and dilation rates are 5 and $\left\{1, 2, 5, 9\right\}$, respectively.

All the utterances are truncated at 6 seconds and sampled at 16 kHz. 20 ms Hanning window is adopted, with 50\% overlap between frames. 320-point FFT is adopted, leading to 161-D features in the frequency dimension. The power spectrum compression technique is adopted to decrease the dynamic range, and the compression factor is empirically set to 0.5~{\cite{li2021importance}}. The model is trained with Adam optimizer~{\cite{kingma2014adam}} and the learning rate (LR) is initialized at 5e-4. We halve the LR if the validation loss does not decrease for consecutive 2 epochs. We train the model for 60 epochs in total and the batch size is 6 at utterance level.

We evaluate the performance of the proposed framework under both causal and non-causal settings. For causal case, three baselines are chosen, namely MMUB~{\cite{ren2021causal}}, EaBNet~{\cite{li2021embedding}}, and frame-wise MVDR beamformer with oracle speech and noise to calculate the covariance matrices~{\cite{higuchi2018frame}}. For non-causal case, four baselines are compared, including FasNet-TAC~{\cite{luo2020end}}, EaBNet, oracle TI-MVDR, and oracle TI-MWF.  
\vspace{-0.3cm}
\section{Results and analysis}
\label{sec:results-and-analysis}
\vspace{-0.2cm}
\subsection{Ablation study}
\label{sec:ablation-study}
\vspace{-0.2cm}
Four objective metrics are adopted, namely PESQ~{\cite{rix2001perceptual}}, ESTOI~{\cite{jensen2016algorithm}}, SISDR~{\cite{le2019sdr}}, and DNSMOS~{\cite{reddy2021dnsmos}}. We conduct ablation studies \emph{w.r.t.} the number of orders $Q$ and microphone channels, whose quantitative results are presented in Figure~{\ref{fig:architecture}}. First, we adjust the value of $Q$ from 0 to 6 to analyze the impact of the Taylor order. It is not surprising to find that with the increase of $Q$, notable performance improvements are achieved for all the metrics, \emph{e.g.}, entries 1a-1f. This is because although the spatial filtering module can suppress most of the directional noises, some residual and diffuse-like interferences still remain. Therefore, with the superimposition of multiple high-order terms, the spectrum can be further refined and improved. Note that with the further increase of $Q$, the performance tends to get plateaued, and the best performance is observed at $Q = 5$.

In entry 2a, the network input only includes the input of the reference channel. In entry 2b, the mixtures from all the channels are sent into the 0th-order module but the signal of the reference microphone is utilized as the input of the high-order encoder module. From entries 2a to 1d, we observe notable improvements among all the metrics, which attest to the importance of spatial information in noise suppression. Besides, going from entries 1d to 2b yields some level of performance degradation, which reveals that despite the spatial filtering operation is applied in both cases, compared to the monaural case, the utilization of spatial information in the high-order modules can still benefit the further cancellation of residual noise. 
\begin{figure}
	\centering
	\subfigure{
		\begin{minipage}[b]{1.00\linewidth}
			\includegraphics[width=\linewidth]{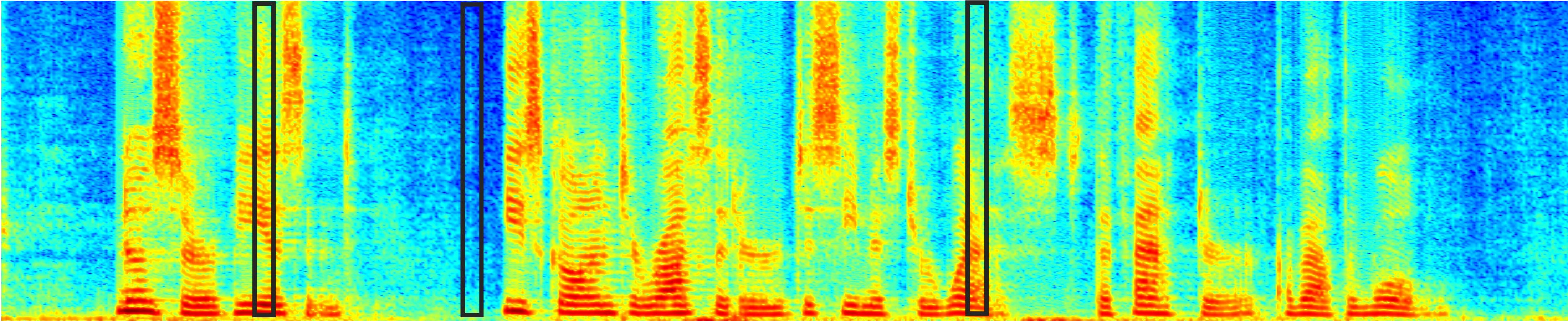}\vspace{-2pt}
			%\centering {\scriptsize Mix}\vspace{-4pt}
	\end{minipage}}
	\subfigure{
		\begin{minipage}[b]{1.00\linewidth}
			\includegraphics[width=\linewidth]{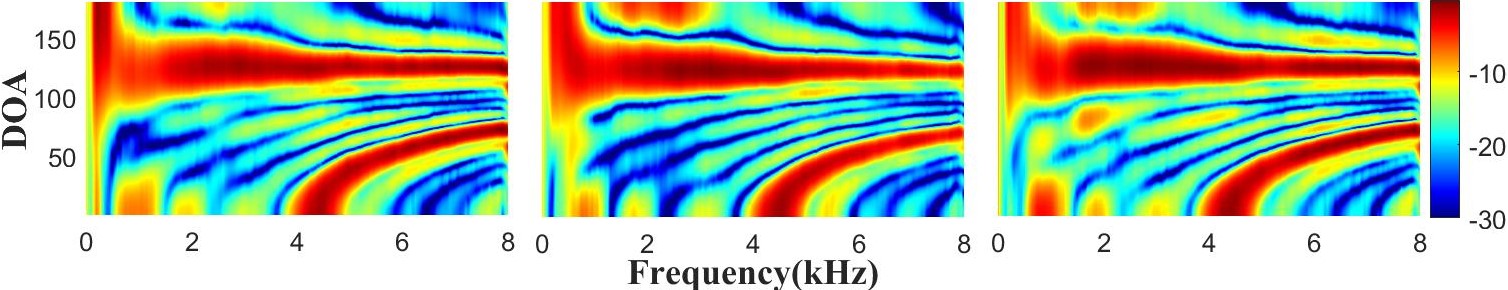}\vspace{-2pt}
			%\centering {\scriptsize $\frac{1}{1!}\mathcal{T}\left(1\right)$}\vspace{-4pt}
	\end{minipage}}
	\vspace{-10pt}
	\caption{A visualization example of the estimated spectrum and the beampatterns at different frame indexes.}
	\vspace{-0.7cm}
	\label{fig:visualization}
\end{figure}
\vspace{-0.3cm}
\subsection{Comparsion with baseline systems}
\label{sec:comparison-with-baseline-systems}
\vspace{-0.2cm}
The network configuration in entry 1d is adopted to compare with other baselines as it well balances between performance and computation complexity. Quantitative results are presented in Table~{\ref{tbl:results-comparison}}. First, not surprisingly, the metric scores of all the systems are consistently improved with the increased difference of DOA between target and noise. This is because a large DOA difference can promote better spatial discriminability between different sources. Second, for causal implmentation, our system significantly outperforms EaBNet, an advanced all-neural frame-level beamformer in the T-F domain, \emph{e.g.}, 2.80/74.75\%/5.91dB/3.25 vs. 2.70/71.92\%/4.81dB/3.20. A similar trend is also observed for non-causal case. It reveals the performance limitation of removing both directional and diffuse-like interferences via only one-step beamforming. In contrast, we can better address them with the collaborative processing of spatial filtering and residual cancellation, which emphasizes the superiority of the proposed Taylor structure. Third, compared with oracle traditional beamformers, the proposed system renders overall better performance in PESQ, ESTOI, and DNSMOS, indicating the advantages of our method in speech quality and subjective quality. Remark that TI-MVDR and TI-MWF are advantageous in SISDR, which may be explained as, under oracle conditions, the linear spatial filter can recover the speech with low distortion. Therefore, it remains to be an unresolved and open problem on how to effectively decrease the speech distortion for all-neural beamformers.

In Figure~{\ref{fig:visualization}}, we present an example of the estimated spectrum and beampatterns at different frame indexes. Target speech and directional noise come from $125^{\circ}$ and $55^{\circ}$, respectively. Obviously, for different frame indexes, the estimated spatial filter weights exhibit a similar beampattern, \emph{i.e.}, it has a main lobe in the target direction while nulling the noise source in the interference direction. Therefore, the 0th-order module indeed works as a spatial filter to extract the target source.
\vspace{-0.4cm}
\section{Conclusions}
\label{sec:conclusion}
\vspace{-0.20cm}
In this paper, we propose an all-neural beamformer called TaylorBeamformer based on Taylor's approximation theory for multi-channel speech enhancement. Concretely, the extraction of target speech is formulated into the spatial filtering of the neighboring point of input mixture and is further decomposed into the superimposition of the 0th-order and multiple high-order derivative terms. For the former, it aims to implement spatial filtering and can suppress most of directional noise. For the latter, trainable modules are employed to simulate the behavior of derivative operations and serve as residual noise canceller for further performance improvement. Experiments on spatialized LibriSpeech corpus well validate the efficacy of the proposed method.
\vfill\pagebreak
\bibliographystyle{IEEEtran}
\bibliography{refs}

\begin{thebibliography}{10}
\providecommand{\url}[1]{#1}
\csname url@samestyle\endcsname
\providecommand{\newblock}{\relax}
\providecommand{\bibinfo}[2]{#2}
\providecommand{\BIBentrySTDinterwordspacing}{\spaceskip=0pt\relax}
\providecommand{\BIBentryALTinterwordstretchfactor}{4}
\providecommand{\BIBentryALTinterwordspacing}{\spaceskip=\fontdimen2\font plus
\BIBentryALTinterwordstretchfactor\fontdimen3\font minus
  \fontdimen4\font\relax}
\providecommand{\BIBforeignlanguage}[2]{{%
\expandafter\ifx\csname l@#1\endcsname\relax
\typeout{** WARNING: IEEEtran.bst: No hyphenation pattern has been}%
\typeout{** loaded for the language `#1'. Using the pattern for}%
\typeout{** the default language instead.}%
\else
\language=\csname l@#1\endcsname
\fi
#2}}
\providecommand{\BIBdecl}{\relax}
\BIBdecl

\bibitem{gannot2017consolidated}
S.~Gannot, E.~Vincent, S.~Markovich-Golan, and A.~Ozerov, ``A consolidated
  perspective on multimicrophone speech enhancement and source separation,''
  \emph{IEEE/ACM Trans.~Audio.~Speech, Lang.~Process.}, vol.~25, no.~4, pp.
  692--730, 2017.

\bibitem{heymann2015blstm}
J.~Heymann, L.~Drude, A.~Chinaev, and R.~Haeb-Umbach, ``Blstm supported {GEV}
  beamformer front-end for the 3rd {CHiME} challenge,'' in
  \emph{Proc.~ASRU}.\hskip 1em plus 0.5em minus 0.4em\relax IEEE, 2015, pp.
  444--451.

\bibitem{qian2018deep}
K.~Qian, Y.~Zhang, S.~Chang, X.~Yang, D.~Florencio, and M.~Hasegawa-Johnson,
  ``Deep learning based speech beamforming,'' in \emph{Proc.~ICASSP}.\hskip 1em
  plus 0.5em minus 0.4em\relax IEEE, 2018, pp. 5389--5393.

\bibitem{erdogan2016improved}
H.~Erdogan, J.~R. Hershey, S.~Watanabe, M.~I. Mandel, and J.~Le~Roux,
  ``Improved mvdr beamforming using single-channel mask prediction networks.''
  in \emph{Proc.~Interspeech}, 2016, pp. 1981--1985.

\bibitem{gu2019neural}
R.~Gu, L.~Chen, S.-X. Zhang, J.~Zheng, Y.~Xu, M.~Yu, D.~Su, Y.~Zou, and D.~Yu,
  ``Neural {S}patial {F}ilter: {T}arget {S}peaker {S}peech {S}eparation
  {A}ssisted with {D}irectional {I}nformation.'' in \emph{Proc.~Interspeech},
  2019, pp. 4290--4294.

\bibitem{wang2018combining}
Z.-Q. Wang and D.~Wang, ``Combining spectral and spatial features for deep
  learning based blind speaker separation,'' \emph{IEEE/ACM
  Trans.~Audio.~Speech, Lang.~Process.}, vol.~27, no.~2, pp. 457--468, 2018.

\bibitem{luo2020end}
Y.~Luo, Z.~Chen, N.~Mesgarani, and T.~Yoshioka, ``End-to-end microphone
  permutation and number invariant multi-channel speech separation,'' in
  \emph{Proc.~ICASSP}.\hskip 1em plus 0.5em minus 0.4em\relax IEEE, 2020, pp.
  6394--6398.

\bibitem{luo2021implicit}
Y.~Luo and N.~Mesgarani, ``Implicit {F}ilter-and-{S}um {N}etwork for
  {E}nd-to-{E}nd {M}ulti-{C}hannel {S}peech {S}eparation,'' in
  \emph{Proc.~Interspeech}, 2021, pp. 3071--3075.

\bibitem{ochiai2020beam}
T.~Ochiai, M.~Delcroix, R.~Ikeshita, K.~Kinoshita, T.~Nakatani, and S.~Araki,
  ``Beam-{T}asnet: Time-domain audio separation network meets frequency-domain
  beamformer,'' in \emph{Proc.~ICASSP}.\hskip 1em plus 0.5em minus 0.4em\relax
  IEEE, 2020, pp. 6384--6388.

\bibitem{zhang2021adl}
Z.~Zhang, Y.~Xu, M.~Yu, S.-X. Zhang, L.~Chen, and D.~Yu, ``{ADL-MVDR}: {A}ll
  deep learning {MVDR} beamformer for target speech separation,'' in
  \emph{Proc.~ICASSP}.\hskip 1em plus 0.5em minus 0.4em\relax IEEE, 2021, pp.
  6089--6093.

\bibitem{halimeh2021complex}
M.~M. Halimeh and W.~Kellermann, ``Complex-valued {S}patial {A}utoencoders for
  {M}ultichannel {S}peech {E}nhancement,'' \emph{arXiv preprint
  arXiv:2108.03130}, 2021.

\bibitem{li2021embedding}
A.~Li, W.~Liu, C.~Zheng, and X.~Li, ``Embedding and {B}eamforming: {A}ll-neural
  {C}ausal {B}eamformer for {M}ultichannel {S}peech {E}nhancement,''
  \emph{arXiv preprint arXiv:2109.00265}, 2021.

\bibitem{casebeer2021nice}
J.~Casebeer, J.~Donley, D.~Wong, B.~Xu, and A.~Kumar, ``{NICE-Beam}: {N}eural
  {I}ntegrated {C}ovariance {E}stimators for {T}ime-{V}arying {B}eamformers,''
  \emph{arXiv preprint arXiv:2112.04613}, 2021.

\bibitem{schwartz2014multi}
O.~Schwartz, S.~Gannot, and E.~A. Habets, ``Multi-microphone speech
  dereverberation and noise reduction using relative early transfer
  functions,'' \emph{IEEE/ACM Trans.~Audio.~Speech, Lang.~Process.}, vol.~23,
  no.~2, pp. 240--251, 2014.

\bibitem{tan2022neural}
K.~Tan, Z.-Q. Wang, and D.~Wang, ``Neural {S}pectrospatial {F}iltering,''
  \emph{IEEE/ACM Trans.~Audio.~Speech, Lang.~Process.}, 2022.

\bibitem{qin2020u2}
X.~Qin, Z.~Zhang, C.~Huang, M.~Dehghan, O.~R. Zaiane, and M.~Jagersand,
  ``U2-net: Going deeper with nested {U}-structure for salient object
  detection,'' \emph{Pattern Recognition}, vol. 106, p. 107404, 2020.

\bibitem{li2021two}
A.~Li, W.~Liu, C.~Zheng, C.~Fan, and X.~Li, ``Two heads are better than one: A
  two-stage complex spectral mapping approach for monaural speech
  enhancement,'' \emph{IEEE/ACM Trans.~Audio.~Speech, Lang.~Process.}, vol.~29,
  pp. 1829--1843, 2021.

\bibitem{panayotov2015librispeech}
V.~Panayotov, G.~Chen, D.~Povey, and S.~Khudanpur, ``Librispeech: an asr corpus
  based on public domain audio books,'' in \emph{Proc.~ICASSP}.\hskip 1em plus
  0.5em minus 0.4em\relax IEEE, 2015, pp. 5206--5210.

\bibitem{allen1979image}
J.~B. Allen and D.~A. Berkley, ``Image method for efficiently simulating
  small-room acoustics,'' \emph{The Journal of the Acoustical Society of
  America}, vol.~65, no.~4, pp. 943--950, 1979.

\bibitem{snyder2015musan}
D.~Snyder, G.~Chen, and D.~Povey, ``Musan: A music, speech, and noise corpus,''
  \emph{arXiv preprint arXiv:1510.08484}, 2015.

\bibitem{li2021importance}
A.~Li, C.~Zheng, R.~Peng, and X.~Li, ``On the importance of power compression
  and phase estimation in monaural speech dereverberation,'' \emph{JASA Express
  Letters}, vol.~1, no.~1, p. 014802, 2021.

\bibitem{kingma2014adam}
D.~P. Kingma and J.~Ba, ``Adam: A method for stochastic optimization,''
  \emph{arXiv preprint arXiv:1412.6980}, 2014.

\bibitem{ren2021causal}
X.~Ren, X.~Zhang, L.~Chen, X.~Zheng, C.~Zhang, L.~Guo, and B.~Yu, ``A {C}ausal
  {U}-net based {N}eural {B}eamforming {N}etwork for {R}eal-{T}ime
  {M}ulti-{C}hannel {S}peech {E}nhancement,'' in \emph{Proc.~Interspeech},
  2021, pp. 1832--1836.

\bibitem{higuchi2018frame}
T.~Higuchi, K.~Kinoshita, N.~Ito, S.~Karita, and T.~Nakatani, ``Frame-by-frame
  closed-form update for mask-based adaptive {MVDR} beamforming,'' in
  \emph{Proc.~ICASSP}.\hskip 1em plus 0.5em minus 0.4em\relax IEEE, 2018, pp.
  531--535.

\bibitem{rix2001perceptual}
A.~W. Rix, J.~G. Beerends, M.~P. Hollier, and A.~P. Hekstra, ``Perceptual
  evaluation of speech quality ({PESQ})-a new method for speech quality
  assessment of telephone networks and codecs,'' in \emph{Proc.~ICASSP},
  vol.~2.\hskip 1em plus 0.5em minus 0.4em\relax IEEE, 2001, pp. 749--752.

\bibitem{jensen2016algorithm}
J.~Jensen and C.~H. Taal, ``An algorithm for predicting the intelligibility of
  speech masked by modulated noise maskers,'' \emph{IEEE/ACM
  Trans.~Audio.~Speech, Lang.~Process.}, vol.~24, no.~11, pp. 2009--2022, 2016.

\bibitem{le2019sdr}
J.~Le~Roux, S.~Wisdom, H.~Erdogan, and J.~R. Hershey, ``Sdr--half-baked or well
  done?'' in \emph{Proc.~ICASSP}.\hskip 1em plus 0.5em minus 0.4em\relax IEEE,
  2019, pp. 626--630.

\bibitem{reddy2021dnsmos}
C.~K. Reddy, V.~Gopal, and R.~Cutler, ``{DNSMOS}: A non-intrusive perceptual
  objective speech quality metric to evaluate noise suppressors,'' in
  \emph{Proc.~ICASSP}.\hskip 1em plus 0.5em minus 0.4em\relax IEEE, 2021, pp.
  6493--6497.

\end{thebibliography}

\end{document}